\documentclass[conference,a4paper]{IEEEtran}

\usepackage{subfig}
\usepackage{float}

\usepackage[cmex10]{amsmath}
\usepackage{amsthm}
\usepackage[mathscr]{eucal}
\usepackage{amssymb}
\usepackage{bbm}

\usepackage{booktabs}

\usepackage{tikz}
\usepackage{pgfplotstable}

\usepackage{hyperref}

\usepackage{enumerate}

\newtheorem{theorem}{Theorem}

\newtheorem{corollary}{Corollary}
\newtheorem{lemma}{Lemma}
\newtheorem{property}{Property}

\newtheorem*{theorem*}{Theorem}
\newtheorem*{exercise*}{Exercise}
\newtheorem*{corollary*}{Corollary}
\newtheorem*{lemma*}{Lemma}
\newtheorem*{property*}{Property}
\newtheorem*{proposition*}{Proposition}
\newtheorem*{problem*}{Problem}
\newtheorem*{observation*}{Observation}

\theoremstyle{definition}

\theoremstyle{remark}
\newtheorem*{remark}{Remark}

\newcommand{\Arikan}{Ar\i kan}

\newcommand{\Ind}[1]{\mathbbm{1}\left[#1\right]}

\newcommand{\Prob}[1]{\mathbb{P} \left[#1\right]}

\newcommand{\E}[1]{\mathbb{E} \left[#1\right]}

\newcommand{\Ebig}[1]{\mathbb{E} \bigl[#1\bigr]}
\newcommand{\EBig}[1]{\mathbb{E} \Bigl[#1\Bigr]}

\DeclareMathOperator{\var}{var}

\newcommand{\varbig}[1]{\var \bigl[#1\bigr]}

\DeclareMathOperator{\cov}{cov}

\newcommand{\covbig}[1]{\cov \bigl[#1\bigr]}
\newcommand{\covBig}[1]{\cov \Bigl[#1\Bigr]}
\newcommand{\covBigg}[1]{\cov \Biggl[#1\Biggr]}

\newcommand{\abs}[1]{\left|#1\right|}
\DeclareMathOperator{\cpop}{CP}
\newcommand{\cp}[1]{\cpop\left[#1\right]}

\newcommand{\bC}{\mathbf{C}}
\newcommand{\bE}{\mathbf{E}}
\newcommand{\bP}{\mathbf{P}}

\newcommand{\bS}{\mathbf{S}}

\newcommand{\bZ}{\mathbf{Z}}

\newcommand{\bp}{\mathbf{p}}
\newcommand{\bs}{\mathbf{s}}
\newcommand{\bt}{\mathbf{t}}

\newcommand{\calA}{\mathcal{A}}
\newcommand{\calC}{\mathcal{C}}
\newcommand{\calD}{\mathcal{D}}
\newcommand{\calE}{\mathcal{E}}
\newcommand{\calF}{\mathcal{F}}
\newcommand{\calG}{\mathcal{G}}

\newcommand{\calS}{\mathcal{S}}

\newcommand{\calX}{\mathcal{X}}
\newcommand{\calY}{\mathcal{Y}}

\newcommand{\barX}{\overline{X}}
\newcommand{\barY}{\overline{Y}}

\newcommand{\barx}{\overline{x}}

\newcommand{\bara}{\overline{a}}
\newcommand{\barb}{\overline{b}}

\newcommand{\e}[2]{ {E_{#1}^{\left(#2\right)}} }
\newcommand{\W}[2]{ {W_{#1}^{\left(#2\right)}} }
\newcommand{\Z}[2]{ {Z_{#1}^{\left(#2\right)}} }
\newcommand{\Zbar}[2]{ {\overline{Z_{#1}^{\left(#2\right)}}} }

\newcommand{\C}[2]{ {C_{#1}^{\left(#2\right)}} }
\newcommand{\Rho}[2]{ {\rho_{#1}^{\left(#2\right)}} }

\newcommand{\BEC}[1]{\mathsf{BEC}\left(#1\right)}

\newcommand{\bigO}[1]{O \left( #1 \right)}
\newcommand{\bigObig}[1]{O \bigl( #1 \bigr)}

\newcommand{\sqfrac}[2]{ \sqrt{\frac{#1}{1+#2}} }

\begin{document}

\sloppy

\title{On the Correlation Between Polarized BECs} 
\author{
  \IEEEauthorblockN{Mani~Bastani~Parizi and Emre~Telatar}
 \IEEEauthorblockA{EPFL, Lausanne, Switzerland\\
   Email: \{mani.bastaniparizi,emre.telatar\}@epfl.ch} %
}
\maketitle
\begin{abstract}
  We consider the $2^n$ channels synthesized by the $n$-fold application of
  \Arikan's polar transform to a binary erasure channel (BEC).  The synthetic
  channels are BECs themselves, and we show that, asymptotically for almost all
  these channels, the pairwise correlations between their erasure events are
  extremely small: the correlation coefficients vanish faster than any
  exponential in $n$.  Such a fast decay of correlations allows us to conclude
  that the union bound on the block error probability of polar codes is very
  tight.  
\end{abstract} 
\section{Introduction} \label{sec:intro}
Channel Polarization is a technique recently introduced by \Arikan{}
\cite{arikan09} as a means of constructing capacity achieving codes for binary
discrete memoryless channels (B-DMCs). The underlying principle of channel
polarization is the following: Let $W: \calX \longrightarrow \calY$ be a B-DMC
with input alphabet $\calX = \mathbb{F}_2$. From two independent copies of $W$
synthesize $W^-: \calX \longrightarrow \calY^2$ and $W^+:\calX \longrightarrow
\calY^2 \times \calX$ as:
\begin{align*} 
   W^-(y_1,y_2 | u_1) &= \sum_{u_2 \in \calX} \frac{1}{2}
   W(y_1|u_1 \oplus u_2) W(y_2|u_2), \\ 
   W^+(y_1,y_2,u_1 | u_2) &= \frac12 W(y_1 | u_1 \oplus u_2) W(y_2|u_2).
\end{align*}
As the superscripts suggest $W^-$ turns out to be a B-DMC worse than $W$ while
$W^+$ is a better B-DMC compared to $W$. 
This transform can be repeated $n$ times to get $N=2^n$ B-DMCs $W_n^{(\bs)}, \bs
\in \{ -,+ \}^n$. \Arikan{} shows that (i) the transformation preserves the
mutual information, (ii) $\W{n}{\bs}$s approach to ``extremal'' channels, i.e.,
either noiseless or useless channels.  In particular, the fraction of almost
noiseless channels is equal to the symmetric capacity of the original B-DMC $W$.
Based on these properties \Arikan{} constructs \emph{polar codes} by sending
uncoded data bits only on (almost) noiseless channels and arbitrary (but known
to receiver) bits on the remaining channels.  The channels used to transmit
information are referred to as ``information'' channels and the rest are called
``frozen'' channels.  A successive cancellation decoder has been proposed by
\Arikan{} to decode the information bits with complexity $\bigO{N \log N}$ and
shown to have a block error probability that behaves roughly as
$\bigObig{2^{-\sqrt{N}}}$ (cf.  \cite{arikan09rate}).

The set of Binary Erasure Channels (BECs) is stable under Polarization in the
sense that if $W$ is a BEC, then $W^+$ and $W^-$ are also BECs.  We denote a
BEC with erasure probability $\epsilon$ as $\BEC{\epsilon}$.  One
can establish a one-to-one relationship between a $\BEC{\epsilon}$ and an
``erasure indicator'' random variable $E$ such that $E \in \{0,1\}$ and $\Prob{E
= 1} = \epsilon$. The polar transform of a BEC is hence equivalent to taking two
independent copies of $E$ and creating the erasure indicators of $W^-$ and
$W^+$. 

\begin{lemma}[Polar Transform of BEC {\cite[Proposition~6]{arikan09}}] 
  \label{lem:bec_transform}
  If $W$ is a BEC with erasure probability $\epsilon$, applying the polar
  transform $(W,W) \mapsto (W^-,W^+)$ produces two BECs $W^+$ with erasure
  probability $\epsilon^2$ and $W^-$ with erasure probability
  $2\epsilon-\epsilon^2$.  Moreover, $W^-$ erases iff either copy of $W$ erases,
  and $W^+$ erases iff both copies of $W$ erase.
\end{lemma}
\begin{corollary} The erasure indicators of $W^-$ and $W^+$, denoted by $E^-$
  and $E^+$, are constructed from two independent copies of $E$, denoted by $E$
  and $E'$, as:
  \begin{subequations}
    \begin{align}
      E^- & = \max\{E,E'\} = E + E' - E E' 	\label{eq:e_minus} \\
      E^+ & = \min\{E,E'\} = E E'. 		\label{eq:e_plus}
    \end{align}
  \end{subequations}
\end{corollary}

While two copies of $E$ are independent (and hence uncorrelated), $E^+$ and
$E^-$ are correlated: $E^+ = 1$ implies $E^- = 1$. On the other side, by
polarization $\W{n}{\bs}$s (and equivalently $\e{n}{\bs}$s) become deterministic
as $n \to \infty$. Hence it looks like $\e{n}{\bs}$ and $\e{n}{\bt}$ would
become uncorrelated for $\bs \ne \bt$, where $\bs$ and $\bt$ are sign sequences
of length $n$ used for indexing the channels.  In particular it is easy to see
that $\Ebig{\e{n}{\bs}\e{n}{\bt}} -\Ebig{\e{n}{\bs}}\Ebig{\e{n}{\bt}}$ is small
for almost every $\bs,\bt$.

In this paper we provide upper bounds on correlation \emph{coefficients} defined
as:
\begin{equation}
  \Rho{n}{\bs,\bt} \triangleq \frac{\Ebig{\e{n}{\bs}\e{n}{\bt}} -
  \Ebig{\e{n}{\bs}}\Ebig{\e{n}{\bt}}} {\sqrt{\varbig{\e{n}{\bs}}
  \varbig{\e{n}{\bt}} } } 			\label{eq:rho_def}
\end{equation}
and exploit these bounds and the inclusion--exclusion principle to find lower
bounds on the block error probability of polar codes.  In particular, our bounds
are strong enough to show that the sum of the Bhattacharyya parameters of the
information channels is a tight estimate of the block error probability.
\section{Notation} Throughout this manuscript, we use uppercase letters (like
$X$) to indicate a random variable, and its lowercase version ($x$) for a
realization of that random variable. The boldface letters denote matrices,
vectors or sequences which will be clear from the context.

We denote the sets by script-style uppercase letters like $\mathcal{S}$ and by
$\abs{\calS}$ we mean the cardinality of $\calS$. 

We use the \emph{bar} notation defined as $\barx \triangleq 1-x$ for the
sake of brevity and refer to $\barx$ as the ``complement'' of $x$.

For sign sequences $\bs \in \{-,+\}^*$ and $\bt \in \{-,+\}^*$, $\cp{\bs,\bt}$
denotes their common prefix.  Furthermore, let $\abs{\bs}$ denote the length of
a sequence $\bs$.
\section{Properties of Correlation Coefficients} \label{sec:basics}
As we mentioned in Section~\ref{sec:intro}, we are interested in analyzing 
the matrix of correlation coefficients of the erasure indicator vector $\bE_n =
\bigl[ \e{n}{\bs}r: \bs \in \{-,+\}^n \bigr]$. It is more convenient to index
the $N=2^n$ elements of that vector using sign sequences $\bs \in \{-,+\}^n$
instead of mapping the sign sequences to integers and using the natural
indexing. We will use the same indexing for the $N^2$ elements of the
correlation coefficients matrix.

\Arikan{} has already shown that the vector $\bZ_n = \Ebig{\bE_n}$ can be
computed via a single-step recursion. More precisely, having $\bZ_{n-1}$ we can
compute the elements of $\bZ_n$ as:
\begin{subequations}
  \begin{align}
    \Z{n}{\bs -} & = 2 \Z{n-1}{\bs} - \left(\Z{n-1}{\bs}\right)^2 	
    \label{eq:z_minus} \\
    \Z{n}{\bs +} & = \left(\Z{n-1}{\bs}\right)^2 			
    \label{eq:z_plus}
  \end{align}
\end{subequations}
for $\forall \bs \in \{-,+\}^{n-1}$ with $Z_0 = \epsilon$.  Note that
\eqref{eq:z_minus} and \eqref{eq:z_plus} can also be derived by taking the
expectation from both sides of \eqref{eq:e_minus} and \eqref{eq:e_plus} and
using the independence between $E$ and $E'$.

Interestingly, the correlation coefficients matrix $\rho_n =
\bigl[\Rho{n}{\bs,\bt}: \bs,\bt \in \{-,+\}^n \bigr]$ can also be computed via a
single-step recursion as we see in this section.

It is useful to rewrite \eqref{eq:e_minus} and \eqref{eq:e_plus} as
\begin{subequations}
  \begin{align}
    E^- &= \overline{ \overline{E} \times \overline{E'}} \\
    E^+ &= E \times E'
  \end{align}
\end{subequations}
and subsequently \eqref{eq:z_minus} and \eqref{eq:z_plus} as:
\begin{subequations}
  \begin{align}
    \Zbar{n}{\bs -} & = \Zbar{n-1}{\bs}^2 	\label{eq:z_minus2} \\
    \Z{n}{\bs +} & = \Z{n-1}{\bs}^2 		\label{eq:z_plus2}
  \end{align}
\end{subequations}
to see the symmetry between `minus' and `plus' transforms.

Recall that the ``covariance'' of random variables $X$ and $Y$ is defined as:
\begin{equation}
  \cov[X,Y] \triangleq \E{X Y} - \E{X} \E{Y}.
\end{equation}

\begin{lemma} \label{lem:cov-bar}
  Let $X$ and $Y$ be arbitrary random variables and set $U \triangleq \barX$ and
  $V \triangleq \barY$. Then:
  \begin{equation}
    \var[U] = \var[X]. \label{eq:var-bar}
  \end{equation}
  Moreover,
  \begin{subequations}
    \begin{align}
      \cov[U,V] & = \cov[X,Y] \label{eq:cov-both-bar} \\
      \cov[X,V] = \cov[U,Y] & = -\cov[X,Y] \label{eq:cov-one-bar}
    \end{align}
  \end{subequations}
\end{lemma}
\begin{IEEEproof} It is clear that $\E{U} = 1 - \E{X}$ and $\E{V} = 1 - \E{Y}$.
  \eqref{eq:var-bar} is also trivial since $\var[aX+b] = \abs{a}^2\var[X]$ for
  any constants $a$ and $b$.  Furthermore:
  \begin{equation*}
    \E{UV} = \E{ (1-X) (1-Y) } = 1 - \E{X} - \E{Y} + \E{XY}
  \end{equation*}
  hence 
  \begin{align*}
    \cov[U,V] & = \E{UV} - \E{U}\E{V} \nonumber \\
    & = \E{XY} - \E{X} \E{Y} = \cov[X,Y]
  \end{align*}
  which proves \eqref{eq:cov-both-bar}. 
  Likewise,
  \begin{equation*}
    \E{UY} = \E{ (1-X) Y } = \E{Y} - \E{XY}
  \end{equation*}
  which shows $\cov[U,Y] = \E{U Y} - \E{U} \E{Y}= -\E{XY} + \E{X}\E{Y} = -
  \cov[X,Y]$. The same argument applies to $\cov[X,V]$ which proves
  \eqref{eq:cov-one-bar}.
\end{IEEEproof}
\begin{corollary} \label{col:rho-bar}
  Let $X$,$Y$,$U$ and $V$ be defined as in Lemma~\ref{lem:cov-bar} and
  $\rho[X,Y] \triangleq \frac{\cov[X,Y]}{\sqrt{\var[X]\var[Y]}}$ denote the
  correlation coefficient between random variables $X$ and $Y$, then:
  \begin{subequations}
    \begin{align}
      \rho[U,V] & = \rho[X,Y] \\
      \rho[X,V] = \rho[U,Y] & = - \rho[X,Y]
    \end{align}
  \end{subequations}
\end{corollary}
\begin{lemma} The covariance matrix of the
  random vector $\bE_n$, $\bC_n \triangleq \bigl[ \C{n}{\bs,\bt}: \bs,\bt \in
  \{-,+\}^n \bigr]$ where
  \begin{equation*} 
    \C{n}{\bs,\bt} \triangleq \covbig{\e{n}{\bs},\e{n}{\bt}}, 
  \end{equation*}
  can be computed in terms of $\bC_{n-1}$ and $\bZ_{n-1}$ as follows:
  \begin{subequations}
    \begin{align}
      \C{n}{\bs-,\bt-} & = 2 \Zbar{n-1}{\bs} \Zbar{n-1}{\bt} \C{n-1}{\bs,\bt} + 
      \C{n-1}{\bs,\bt}^2, \label{eq:c_mm} \\
      \C{n}{\bs-,\bt+} & = 2 \Zbar{n-1}{\bs} \Z{n-1}{\bt} \C{n-1}{\bs,\bt} - 
      \C{n-1}{\bs,\bt}^2, \label{eq:c_mp} \\
      \C{n}{\bs+,\bt-} & = 2 \Z{n-1}{\bs} \Zbar{n-1}{\bt} \C{n-1}{\bs,\bt} - 
      \C{n-1}{\bs,\bt}^2, \label{eq:c_pm} \\
      \C{n}{\bs+,\bt+} &= 2 \Z{n-1}{\bs} \Z{n-1}{\bt} \C{n-1}{\bs,\bt} + 
      \C{n-1}{\bs,\bt}^2. \label{eq:c_pp}
    \end{align}
  \end{subequations}
  It is clear that $C_0 = \epsilon \overline{\epsilon}$ where $\epsilon$ is the
  erasure probability of the underlying BEC.
\end{lemma}
\begin{IEEEproof} We first prove \eqref{eq:c_pp} and then show how the rest of
  results easily follow using Lemma~\ref{lem:cov-bar}.

  Recall that $\e{n}{\bs+} = \e{n-1}{\bs} \times \e{n-1}{\bs}'$ and $\e{n}{\bt+}
  = \e{n-1}{\bt} \times \e{n-1}{\bt}'$. Furthermore, $\Ebig{\e{n-1}{\bs}} =
  \Z{n-1}{\bs}$ and $\Ebig{\e{n-1}{\bt}} = \Z{n-1}{\bt}$:
  \begin{align*}
    & \covbig{\e{n}{\bs+}, \e{n}{\bt+}}  = \EBig{\e{n-1}{\bs} \e{n-1}{\bs}'
    \e{n-1}{\bt} \e{n-1}{\bt}'} \\ 
    & \qquad -  \EBig{\e{n-1}{\bs} \e{n-1}{\bs}'}
    \EBig{\e{n-1}{\bt} \e{n-1}{\bt}'} \\
    & \stackrel{(*)}{=} \Ebig{\e{n-1}{\bs} \e{n-1}{\bt}}^2 -
    \Ebig{\e{n-1}{\bs}}^2 \Ebig{\e{n-1}{\bt}}^2 \\
    & = \Bigl( \Ebig{\e{n-1}{\bs} \e{n-1}{\bt}} - \Z{n-1}{\bs} \Z{n-1}{\bt}
    \Bigr)^2 \\
    & \qquad + 2 \Z{n-1}{\bs} \Z{n-1}{\bt} \Bigl( \Ebig{\e{n-1}{\bs}
    \e{n-1}{\bt}} 
    - \Z{n-1}{\bs} \Z{n-1}{\bt} \Bigr) \\
    & = \C{n-1}{\bs,\bt}^2 + 2 \Z{n-1}{\bs} \Z{n-1}{\bt} \C{n-1}{\bs,\bt}.
  \end{align*}
  Note that in (*) we have used the independence between the indicator variables
  with \emph{prime} and the ones without that and the fact that they are both
  identical copies of the same random variable.

  Now observe that $\e{n}{\bs-} = \overline{ \overline{\e{n-1}{\bs}} \times
  \overline{\e{n-1}{\bs}'}}$ and $\e{n}{\bt-} =
  \overline{\overline{\e{n-1}{\bt}} \times \overline{\e{n-1}{\bt}'}}$. 
  
  To compute $\C{n}{\bs-,\bt-}$, using \eqref{eq:cov-both-bar} we have:
  \begin{align*}
    \covbig{\e{n}{\bs-}, \e{n}{\bt-}} & = \covBigg{\overline{ \overline{\e{n-1}{\bs}} \times
    \overline{\e{n-1}{\bs}'}},\overline{ \overline{\e{n-1}{\bt}} \times
    \overline{\e{n-1}{\bt}'}}} \\
    & = \covBig{\overline{\e{n-1}{\bs}} \times \overline{\e{n-1}{\bs}'},
    \overline{\e{n-1}{\bt}} \times \overline{\e{n-1}{\bt}'}} \\
    & \stackrel{(*)}{=} \C{n-1}{\bs,\bt}^2 + 2 \Zbar{n-1}{\bs} \Zbar{n-1}{\bt}
    \C{n-1}{\bs,\bt} 
  \end{align*}
  where (*) follows by observing that we are essentially computing the same
  covariance as the one we just computed to show \eqref{eq:c_pp} considering the
  facts that (i) $\covBig{\overline{\e{n-1}{\bs}},\overline{\e{n-1}{\bt}}} =
  \covbig{\e{n-1}{\bs},\e{n-1}{\bt}}$ (using \eqref{eq:cov-both-bar} once again)
  and (ii) $\EBig{\overline{\e{n-1}{\bs}}} = \Zbar{n-1}{\bs}$ and
  $\EBig{\overline{\e{n-1}{\bt}}} = \Zbar{n-1}{\bt}$.

  Likewise \eqref{eq:c_mp} (similarly \eqref{eq:c_pm}) follows using
  \eqref{eq:cov-one-bar}:
  \begin{align*}
    \covbig{\e{n}{\bs-},\e{n}{\bt+}} & = \covBigg{\overline{ \overline{\e{n-1}{\bs}}
    \times \overline{\e{n-1}{\bs}'}}, \e{n-1}{\bt} \times \e{n-1}{\bt}'} \\ 
    & = - \covBig{\overline{\e{n-1}{\bs}} \times \overline{\e{n-1}{\bs}'},
    \e{n-1}{\bt} \times \e{n-1}{\bt}'} \\ 
    & \stackrel{(*)}{=} - \bigl( \C{n-1}{\bs,\bt}^2 - 2 \Zbar{n-1}{\bs}
    \Z{n-1}{\bt} \C{n-1}{\bs,\bt} \bigr).
  \end{align*}
  Once again in (*) we are computing the same form of covariance as the one we
  did to show \eqref{eq:c_pp} considering the fact that
  $\covBig{\overline{\e{n-1}{\bs}}, \e{n-1}{\bt}} = -
  \covbig{\e{n-1}{\bs},\e{n-1}{\bt}} = - \C{n-1}{\bs,\bt}$ (by
  \eqref{eq:cov-one-bar}).  
\end{IEEEproof}
\begin{corollary}  \label{col:rho_computations}
  The correlation coefficients matrix of the random vector $\bE_n$, defined as
  $\mathbf{\rho}_n \triangleq \Bigl[ \Rho{n}{\bs,\bt} \Bigr]$ (where
  $\Rho{n}{\bs,\bt}$ is defined in \eqref{eq:rho_def}) can be computed
  in terms of $\rho_{n-1}$ and $\bZ_{n-1}$ as:
  \begin{subequations}
    \begin{align}
      \Rho{n}{\bs-,\bt-} &= 2 \sqfrac{\Zbar{n-1}{\bs}}{\Zbar{n-1}{\bs}}
      \sqfrac{\Zbar{n-1}{\bt}}{\Zbar{n-1}{\bt}} \Rho{n-1}{\bs,\bt} \nonumber \\
      & \qquad +
      \sqfrac{\Z{n-1}{\bs}}{\Zbar{n-1}{\bs}}
      \sqfrac{\Z{n-1}{\bt}}{\Zbar{n-1}{\bt}} \Rho{n-1}{\bs,\bt}^2
      \label{eq:rho_mm} \\
      \Rho{n}{\bs-,\bt+} &= 2 \sqfrac{\Zbar{n-1}{\bs}}{\Zbar{n-1}{\bs}}
      \sqfrac{\Z{n-1}{\bt}}{\Z{n-1}{\bt}} \Rho{n-1}{\bs,\bt} \nonumber \\ 
      & \qquad -
      \sqfrac{\Z{n-1}{\bs}}{\Zbar{n-1}{\bs}}
      \sqfrac{\Zbar{n-1}{\bt}}{\Z{n-1}{\bt}} \Rho{n-1}{\bs,\bt}^2
      \label{eq:rho_mp} \\
      \Rho{n}{\bs+,\bt-} &= 2 \sqfrac{\Z{n-1}{\bs}}{\Z{n-1}{\bs}} 
      \sqfrac{\Zbar{n-1}{\bt}}{\Zbar{n-1}{\bt}} \Rho{n-1}{\bs,\bt} \nonumber \\ 
      & \qquad -
      \sqfrac{\Zbar{n-1}{\bs}}{\Z{n-1}{\bs}} \sqfrac{\Z{n-1}{\bt}}{\Zbar{n-1}{\bt}}
      \Rho{n-1}{\bs,\bt}^2    \label{eq:rho_pm} \\
      \Rho{n}{\bs+,\bt+} &= 2 \sqfrac{\Z{n-1}{\bs}}{\Z{n-1}{\bs}} 
      \sqfrac{\Z{n-1}{\bt}}{\Z{n-1}{\bt}} \Rho{n-1}{\bs,\bt} \nonumber \\
      & \qquad + 
      \sqfrac{\Zbar{n-1}{\bs}}{\Z{n-1}{\bs}} \sqfrac{\Zbar{n-1}{\bt}}{\Z{n-1}{\bt}}
      \Rho{n-1}{\bs,\bt}^2    \label{eq:rho_pp}
    \end{align}
  \end{subequations}
  Clearly $\rho_0 = 1$.
\end{corollary}
\begin{IEEEproof} Once again we only prove \eqref{eq:rho_pp} and the rest follow
  by the symmetry using Corollary~\ref{col:rho-bar}.
  Since $\e{n}{\bs}$s are $\{0,1\}$ valued RVs with mean $\Z{n}{\bs}$:
  \begin{equation}
    \var[\e{n}{\bs}]  = \Z{n}{\bs} \Zbar{n}{\bs}.
    \label{eq:var}
  \end{equation}
  Setting $\C{n}{\bs,\bt} = \Rho{n}{\bs,\bt}\sqrt{\Z{n}{\bs}\Zbar{n}{\bs}
  \Z{n}{\bt}\Zbar{n}{\bt}}$ in both sides of \eqref{eq:c_pp} and using the fact
  that $\Z{n}{\bs+} = \Z{n-1}{\bs}^2$ (similarly $\Z{n}{\bt+} = \Z{n-1}{\bt}^2$)
  we get:
  \begin{align*}
    & \Rho{n}{\bs+,\bt+} \sqrt{\Z{n-1}{\bs}^2 
    \overline{\left(\Z{n-1}{\bs}^2\right)} \Z{n-1}{\bt}^2 
    \overline{\left(\Z{n-1}{\bt}^2\right)}} = \\ 
    & \qquad 2 \Z{n-1}{\bs} \Z{n-1}{\bt} \sqrt{ \Z{n-1}{\bs} \Zbar{n-1}{\bs} \Z{n-1}{\bt}
    \Zbar{n-1}{\bt}} \Rho{n-1}{\bs,\bt} \\
    & \qquad  + \bigl(\Z{n-1}{\bs}\Zbar{n-1}{\bs} \Z{n-1}{\bt} 
    \Zbar{n-1}{\bt}\bigr) \Rho{n-1}{\bs,\bt}^2 
  \end{align*}
  Eliminating $\Z{n-1}{\bs} \Z{n-1}{\bt}$ from both sides and observing
  that $\sqrt{\frac{x \barx}{1-x^2}} = \sqfrac{x}{x}$ and
  $\frac{\barx}{\sqrt{1-x^2}} = \sqfrac{\barx}{x}$ proves the claim.
\end{IEEEproof}

The property of being computable by a single-step recursion generalizes to
higher order statistics:

\begin{lemma} In general the $m$-th order moments of the random variables
  $\e{n}{\bs^n}, \bs^n \in \{-,+\}^n$ can be computed from the $m$-th order
  moments of random variables $\e{n-1}{\bs^{n-1}}, \bs^{n-1} \in \{-,+\}^{n-1}
  $.  
\end{lemma}
\begin{IEEEproof} By the $m$-th order moment we mean:
  \begin{equation*}
    \EBig{\e{n}{\bs_1^{n}} \e{n}{\bs_2^{n}} \cdots  \e{n}{\bs_m^{n}}} 
  \end{equation*}
  for some set of indices $\bs_1^{n},\bs_2^{n},\cdots,\bs_m^{n}$ which are
  \emph{not} necessarily distinct.

  Let $\bs^{n-1}$ denote the subsequence of $\bs^n$ including its first $n-1$
  elements and observe that for any $k \in \{1,2,\dots,m\}$, $\e{n}{\bs_k^n}$ is
  linear in each of $\e{n-1}{\bs^{n-1}_k}$ and $\e{n-1}{\bs^{n-1}_k}'$
  (cf.~\eqref{eq:e_minus} and \eqref{eq:e_plus}). This means in the expansion of
  $\e{n}{\bs^n_1} \e{n}{\bs^n_2} \cdots \e{n}{\bs^n_m}$ we will have the terms
  in the form of $\e{n-1}{\bs^{n-1}_1} \e{n-1}{\bs^{n-1}_2} \cdots
  \e{n-1}{\bs^{n-1}_l} \times \e{n-1}{ {\bs^{n-1}_1}'}'\e{n-1}{ {\bs^{n-1}_2}'}'
  \cdots \e{n-1}{ {\bs^{n-1}_{l'}}'}'$ for some $l \le m$ and $l' \le m$. 

  The independence of the variables with prime and the one without prime implies
  that the expectation of such product will be product of two expectations
  each of which is at most an $m$-th order moment of the random variables
  $\e{n-1}{\bs^{n-1}}$.
\end{IEEEproof}

One can derive the properties stated in the sequel on $\Rho{n}{\bs,\bt}$
according to the aforementioned recursions: 
\begin{property} \label{prop:rho-bound}
  \begin{equation}
    0 \le \Rho{n}{\bs,\bt} \le \min \left\{ \sqrt{\frac
      {\Zbar{n}{\bs}\Z{n}{\bt}} {\Z{n}{\bs} \Zbar{n}{\bt}}}, \sqrt{
	\frac{\Z{n}{\bs} \Zbar{n}{\bt}}{\Zbar{n}{\bs}\Z{n}{\bt}} } \right\}
  \end{equation}
\end{property}

Property~\ref{prop:rho-bound} follows as a corollary of the following property
on $\C{n}{\bs,\bt}$:
\begin{property*}
  \begin{equation}
    0 \le \C{n}{\bs,\bt} \le
    \min \Bigl\{\Zbar{n}{\bs}\Z{n}{\bt},\Z{n}{\bs}\Zbar{n}{\bt} \Bigr\}
    \label{eq:c-bound}
  \end{equation}
\end{property*}
\begin{IEEEproof}  We prove the claim by induction on $n$.  
  The claim is trivially true for $n = 0$ since:
  \begin{equation*}
    0 \le C_{0} = \var[E_{0}] = \epsilon \overline{\epsilon} \le
    \min \{ \overline{\epsilon} \epsilon, \epsilon \overline{\epsilon} \}
  \end{equation*}
  where $\epsilon$ is the erasure probability of the underlying BEC. 

  Now, assuming \eqref{eq:c-bound} holds for $n-1$, we shall show:
  \begin{subequations}
    \begin{align}
      0 \le \C{n}{\bs-,\bt-} & \le \min \Bigl\{ \Zbar{n}{\bs-} \Z{n}{\bt-},
      \Z{n}{\bs-} \Zbar{n}{\bt-} \Bigr\}. \label{eq:bound_mm} \\
      0 \le \C{n}{\bs-,\bt+} & \le \min \Bigl\{ \Zbar{n}{\bs-} \Z{n}{\bt+},
      \Z{n}{\bs-} \Zbar{n}{\bt+} \Bigr\}.  \label{eq:bound_mp}  \\
      0 \le \C{n}{\bs+,\bt-} & \le \min \Bigl\{ \Zbar{n}{\bs+} \Z{n}{\bt-}, 
      \Z{n}{\bs+} \Zbar{n}{\bt-} \Bigr\}. \label{eq:bound_pm}  \\
      0 \le \C{n}{\bs+,\bt+} & \le \min \Bigl\{ \Zbar{n}{\bs+} \Z{n}{\bt+}, 
      \Z{n}{\bs+} \Zbar{n}{\bt+} \Bigr\}. \label{eq:bound_pp}  
    \end{align}
  \end{subequations}

  As \eqref{eq:c_pp} is obtained by replacing both $\Z{n}{\bt}$ and $\Z{n}{\bs}$
  by their complements and \eqref{eq:c_pm} is obtained by swapping $\bs$ and
  $\bt$ in \eqref{eq:c_mp} we only need to prove \eqref{eq:bound_mm} and
  \eqref{eq:bound_mp} and the rest follow by symmetry. Furthermore, positivity
  of $\C{n}{\bs-,\bt-}$ and $\C{n}{\bs-,\bt+}$ is clear by the assumption
  \eqref{eq:c-bound} (for $n-1$) and the combination formulae \eqref{eq:c_mm}
  and \eqref{eq:c_mp}. So, we only verify the upper-bounds.
 
  Let $a \triangleq \Z{n-1}{\bs}$, $b \triangleq \Z{n-1}{\bt}$ and $c \triangleq
  \C{n-1}{\bs,\bt}$, for the sake of brevity.  Note that by definition $0 \le a
  \le 1$ and $0 \le b \le 1$.  However, if either $a$ or $b$ is extremal, by
  assumption \eqref{eq:c-bound}, $c =0$ and the claim is trivial. So, for the
  rest of the proof, we safely assume $0 < a <1$ and $0 < b <1$.  
  \begin{itemize}
    \item To prove \eqref{eq:bound_mm} we have to show:
      \begin{equation*}
	2 \bara \barb c + c^2  \le \min \{ \bara^2 (2b-b^2),
	(2a - a^2) \barb^2 \}. 
      \end{equation*}
      The above inequality is symmetric in $a$ and $b$ hence without loss of
      generality we can assume $a \ge b$ which implies $\barb a \ge \bara b$ and
      also $(2a - a^2) \barb^2 \ge \bara^2 (2b - b^2)$. 
      The LHS of the above inequality is increasing in $c$, hence once we verify
      the inequality for maximum possible value of $c$ we are done. Replacing
      $c$ with $\bara b$ we get: 
      \begin{equation*}
	2 \bara \barb \bara b  + (\bara b)^2  \le (\bara)^2 (2b - b^2).
      \end{equation*}
      Simplifying $\bara^2$ from both sides yields $2b - b^2 \le 2b
      -b^2$.
    \item To prove \eqref{eq:bound_mp} we need to show:
      \begin{align*}
	2 \bara b c - c^2 & \le \min \{ \bara^2 b^2, (2a-a^2)(1-b^2) \} \\
	 & = \min \{(\bara b)^2, a \barb (1+\bara) (1+b)\}
      \end{align*}
      As $c \le \bara b$ the LHS is an increasing function of $c$ and we only
      need to verify the inequality for maximum possible value of $c$.
      \begin{itemize}
	\item If $\bara b \le a \barb$, the LHS of the inequality will be
	  $(\bara b)^2$ at $c = \bara b$ and:
	  \begin{equation*}
	    (\bara b)^2 \le (\bara b) \times (a \barb) <
	    \left[(1+\bara)(1+b)\right] \times \left[a \barb \right]
	  \end{equation*}
	\item If $a \barb \le \bara b$, then the LHS of
	  our inequality at $c = a \barb$ will be equal to:
	  \begin{align*}
	    2 \bara b \times a \barb - \left(a \barb\right)^2 & = a \barb \left[
	      2 \bara b - a \barb \right] \\
	    & = a \barb \left[ \bara b + \bara + b  - 1 \right] \\
	    & = a \barb \left[ (1 + \bara) (1+b) - 2 \right]  \\
	    & \le a \barb (1+\bara) (1+b)
	  \end{align*}
	  Furthermore, as the LHS is increasing in $c$, at $c = a
	  \barb$ it will be less than $(\bara b)^2$ (its value at $c =
	  \bara b$). \hfill \IEEEQED \\[-5ex]
      \end{itemize}
  \end{itemize}
  \let\IEEEQED\relax 
\end{IEEEproof}
\begin{remark} This upper-bound shows for almost all choices of
  $\bs$ and $\bt$, $\C{n}{\bs,\bt} = \Ebig{\e{n}{\bs}\e{n}{\bt}} -
  \Ebig{\e{n}{\bs}}\Ebig{\e{n}{\bt}}$ goes to zero as $n$ gets large.
\end{remark}
\begin{property} \label{prop:decreasing} 
  For $\bs,\bt \in \{-,+\}^{n-1}$ and $s_n,t_n \in \{-,+\}$ 
  \begin{equation*}
   \Rho{n}{\bs s_n, \bt t_n} \le \Rho{n-1}{\bs,\bt} 
  \end{equation*}
  with equality iff 
  \begin{enumerate}[(i)]
    \item $\Rho{n-1}{\bs,\bt} = 0$, or
    \item $s_n = t_n$ and $\Rho{n-1}{\bs,\bt} = 1$ and $\Z{n-1}{\bs} =
      \Z{n-1}{\bt}$, or
    \item $\Z{n-1}{\bs}=b_{s_n}$ and $\Z{n-1}{\bt}=b_{t_n}$, where $b_+=1$ and
      $b_-=0$.
  \end{enumerate}
\end{property}
\begin{IEEEproof} The case of $\Rho{n-1}{\bs,\bt} = 0$ is trivial. Otherwise, we
  consider the ratio ${\Rho{n}{\bs s_n,\bt t_n}}/
  {\Rho{n-1}{\bs,\bt}}$. Using \eqref{eq:rho_mm} to \eqref{eq:rho_pp} this ratio
  is as shown in \eqref{eq:ratio_cases}.
  \begin{figure*}[!tb]
    \normalsize
    \begin{equation}
      \frac{\Rho{n}{\bs s_n,\bt t_n}}{\Rho{n-1}{\bs,\bt}} = \begin{cases}
	2 \sqfrac{\Z{n-1}{\bs}}{\Z{n-1}{\bs}}
	\sqfrac{\Z{n-1}{\bt}}{\Z{n-1}{\bt}} +
	\sqfrac{\Zbar{n-1}{\bs}}{\Z{n-1}{\bs}}
	\sqfrac{\Zbar{n-1}{\bt}}{\Z{n-1}{\bt}} \Rho{n-1}{\bs,\bt}  & \text{if }
	(s_n,t_n) = (+,+), \\
	2 \sqfrac{\Z{n-1}{\bs}}{\Z{n-1}{\bs}}
	\sqfrac{\Zbar{n-1}{\bt}}{\Zbar{n-1}{\bt}} -
	\sqfrac{\Zbar{n-1}{\bs}}{\Z{n-1}{\bs}}
	\sqfrac{\Z{n-1}{\bt}}{\Zbar{n-1}{\bt}} \Rho{n-1}{\bs,\bt}  & \text{if }
	(s_n,t_n) = (+,-), \\
	2 \sqfrac{\Zbar{n-1}{\bs}}{\Zbar{n-1}{\bs}}
	\sqfrac{\Z{n-1}{\bt}}{\Z{n-1}{\bt}} -
	\sqfrac{\Z{n-1}{\bs}}{\Zbar{n-1}{\bs}}
	\sqfrac{\Zbar{n-1}{\bt}}{\Z{n-1}{\bt}} \Rho{n-1}{\bs,\bt}  & \text{if }
	(s_n,t_n) = (-,+), \\
	2
	\sqfrac{\Zbar{n-1}{\bs}}{\Zbar{n-1}{\bs}}\sqfrac{\Zbar{n-1}{\bt}}{\Zbar{n-1}{\bt}}
	+ \sqfrac{\Z{n-1}{\bs}}{\Zbar{n-1}{\bs}}
	\sqfrac{\Z{n-1}{\bt}}{\Zbar{n-1}{\bt}} \Rho{n-1}{\bs,\bt}  & \text{if }
	(s_n,t_n) = (-,-). 
      \end{cases}
      \label{eq:ratio_cases}
    \end{equation}
    \hrulefill
  \end{figure*}
  Let $a \triangleq \Z{n-1}{\bs}$, $b \triangleq \Z{n-1}{\bt}$ and $r \triangleq
  \Rho{n-1}{\bs,\bt}$ and observe that:
  \begin{enumerate}
    \item If $(s_n,t_n) = (+,+)$, applying the Cauchy-Schwarz inequality to the
      RHS of \eqref{eq:ratio_cases} we get:
      \begin{align*}
	& \left( \sqfrac{2a}{a} \sqfrac{2b}{b} + \sqfrac{\bara r}{a}
	\sqfrac{\barb r}{b} \right) \\ & \qquad \le \sqfrac{2a +
	r \bara}{a} \sqfrac{2b + r\barb}{b}
      \end{align*}
      For $a \in [0,1]$, $b \in [0,1]$ and $r \in [0,1]$, each of the
      square-roots are strictly smaller than $1$ unless $r = 1$ \footnote{As
	each of them is in the form of $\sqfrac{1+x + \left(r - 1\right)
	\barx}{x}$ which is smaller than one since the numerator is less than
      the denominator.} or $a = b = 1$. Furthermore, the equality conditions for
      Cauchy-Schwarz inequality imply $\sqrt{a / \bara} = \sqrt{b / \barb}$
      which in turn implies $a = b$.  Therefore, we can conclude that if $(s_n,
      t_n) = (+,+)$, ${\Rho{n-1}{\bs s_n,\bt t_n}}/{\Rho{n-1}{\bs,\bt}} \le
      1$ with equality iff ($\Z{n-1}{\bs} = \Z{n-1}{\bt}$ and
      $\Rho{n-1}{\bs,\bt} = 1$) or ($\Z{n-1}{\bs} = \Z{n-1}{\bt} = 1$).

      The same argument can also be applied to the case of $(s_n,t_n) = (-,-)$.
    \item If $(s_n,t_n)=(+,-)$, the RHS of \eqref{eq:ratio_cases} can be
      bounded as:
     \begin{align*}
	& {2\sqfrac{a}{a}\sqfrac{\barb}{\barb} -
	\sqfrac{\bara}{a}\sqfrac{b}{\barb} r} \\ & \qquad \le 2 \sqfrac{a}{a}
	\sqfrac{\barb}{\barb} \le 1.
      \end{align*}
      The last inequality follows by observing that $\sqfrac{x}{x} \le
      \frac{1}{\sqrt2}$ for $x \in [0,1]$ with equality iff $x = 1$.
      Furthermore, it is easy to see that the equality in all obove chain of
      weak inequalities happens iff $(a,b) = (1,0)$\footnote{By
	Property~\ref{prop:rho-bound} this condition implies
	$\Rho{n-1}{\bs,\bt} = 0$.}.
      By symmetry, this argument also applies to the case of $(s_n,t_n)= (-,+)$.
      \hfill \IEEEQED \\[-5ex]
  \end{enumerate}
  \let\IEEEQED\relax
\end{IEEEproof}

\begin{property} \label{prop:one-third}
  If $\bs \ne \bt$ then $\Rho{n}{\bs,\bt} \le \frac{1}{3}$.
\end{property}
\begin{IEEEproof} Let $\bp \triangleq \cp{\bs,\bt}$ be the common prefix of
  $\bs$ and $\bt$ and $m \triangleq \abs{\bp}$ its length. Then $s_{m+1} \ne
  t_{m+1}$ and Property~\ref{prop:decreasing} together with either
  \eqref{eq:rho_mp} or \eqref{eq:rho_pm} result in:
  \begin{align*}
    \Rho{n}{\bs,\bt} & \le \Rho{m+1}{\bp s_{m+1},\bp t_{m+1}} \\
    & = 2 \sqfrac{\Z{m}{\bp}}{\Z{m}{\bp}} 
    \sqfrac{\Zbar{m}{\bp}}{\Zbar{m}{\bp}} - 
    \sqfrac{\Zbar{m}{\bp}}{\Z{m}{\bp}} \sqfrac{\Z{m}{\bp}}{\Zbar{m}{\bp}} \\
    & = \sqrt{ \frac{\Z{m}{\bp} \Zbar{m}{\bp}}
    { \left(1+\Z{m}{\bp}\right) \left(1+\Zbar{m}{\bp}\right) } } 
    = \sqrt{ \frac{\Z{m}{\bp} \Zbar{m}{\bp}}
    {2 + \Z{m}{\bp} \Zbar{m}{\bp}}}  \le \frac{1}{3}
  \end{align*}
  with equality iff $\Z{m}{\bp} = \frac{1}{2}$.
\end{IEEEproof}
\section{Convergence of Correlation Coefficients} In the previous section we
showed how correlation coefficients can be computed efficiently by single-step
recursions and derived some algebraic properties of them. In this section we
show that correlation coefficients converge to zero. 

\begin{lemma}\label{lem:sample-wise-decrease}
  Let $\bs$ and $\bt$ be infinite sign sequences such that $\bs \ne \bt$ and
  $\bs^n$ and $\bt^n$ be the subsequences corresponding to their first $n$
  elements  respectively. Then $\lim_{n\to \infty} \Rho{n}{\bs^n,\bt^n} = 0$. 
\end{lemma}
\begin{IEEEproof} 
  Let $m = \abs{\cp{\bs,\bt}}$ and $a_n \triangleq \Rho{n}{\bs^n,\bt^n}$. For $n
  > m$, by Properties~\ref{prop:rho-bound} and \ref{prop:one-third} we know $a_n
  \in [0,1/3]$ and by Property~\ref{prop:decreasing} it is decreasing. Hence,
  $a_n$ is a convergent sequence. Suppose its limit is $a^* > 0$. This implies
  for every $\varepsilon > 0$ there exist a $n_0$ such that for $n > n_0$,
  $a_n/a_{n-1}  \ge 1 - \varepsilon$.  By the continuity
  of~\eqref{eq:ratio_cases}, we must have $|\Z{n-1}{\bs^{n-1}} -
  b_{s_n}|<\delta$ and $|\Z{n-1}{\bt^{n-1}} - b_{t_{n}}|<\delta$ for all $n >
  n_0$ according to equality condition (iii) of Property~\ref{prop:decreasing},
  where $\delta$ is a quantity approaching zero as $\varepsilon$ gets small.
  This implies $s_n = s^*$ and $t_n = t^*$ for all $n > n_0$ because the
  evolutions of $Z$ do not allow $Z$ to jump from one extreme to the other.
  Without loss of generality, assume $s^* = +$ which in turn requires
  $\Z{n-1}{\bs^{n-1}} > 1 - \delta$. Now we have an incompatible situation:
  $s_n=+$ for all $n>n_0$ will drive $\Z{n}{\bs^n}$ to 0.  This shows $a_n$
  cannot converge to a non-zero value. 
\end{IEEEproof}

Additionally we can show that the average of the elements of the correlation
coefficients matrix is exponentially small in $n$.
\begin{lemma} \label{lem:rho-average}
  For any $\bs, \bt \in \{-,+\}^{n-1}$, 
  \begin{equation*}
    \frac{1}{4} \sum_{(s,t) \in \{-,+\}^2} \Rho{n}{\bs s,\bt t} \le \frac{2}{3}
    \Rho{n-1}{\bs,\bt}.
  \end{equation*}
\end{lemma}
\begin{IEEEproof}
  Let $a = \Z{n-1}{\bs}$, $b = \Z{n-1}{\bt}$, $f(x) \triangleq \allowbreak
  \frac{1}{\sqrt2} \Bigl[ \sqfrac{x}{x} + \sqfrac{\barx}{\barx} \Bigr]$, and
  $g(x) \triangleq \allowbreak \frac12 \Bigl[ \sqfrac{\barx}{x} -
    \sqfrac{x}{\barx} \Bigr]$. Using \eqref{eq:rho_mm} to \eqref{eq:rho_pp} one
  can easily verify that:
  \begin{align*}
    \frac{1}{4} \sum_{(s,t) \in \{-,+\}^2} \Rho{n}{\bs s,\bt t}  & =
    f(a) f(b) \Rho{n-1}{\bs,\bt} + g(a) g(b) \Rho{n-1}{\bs,\bt}^2  \\
     & = \Bigl[ f(a)f(b) +  g(a) g(b) \Rho{n-1}{\bs,\bt} \Bigr]
    \Rho{n-1}{\bs,\bt}.
  \end{align*} 
  Now, observe that both sides of the above are positive and:
  \begin{align*}
     & \left[f(a)f(b) + g(a) g(b) \Rho{n-1}{\bs,\bt}\right]^2  \\
    & \quad \stackrel{\text{(*)}}{\le} \left[ f(a)^2 + \Rho{n-1}{\bs,\bt}
      g(a)^2 \right] \left[ f(b)^2 + \Rho{n-1}{\bs,\bt} g(b)^2 \right] \\ 
    & \quad \le \left[ f(a)^2 + g(a)^2 \right] \left[ f(b)^2 + g(b)^2 \right] 
  \end{align*}
  where (*) follows from the Cauchy-Schwarz inequality. It is easy to see
  $f(x)^2+g(x)^2 = \frac{1}{2} \left(1 + \sqrt{ \frac{x \barx}{(1+x)(1+\barx)}}
  \right)$ which is maximized at $x = \frac{1}{2}$ (for $x \in [0,1]$) with
  value $\frac{2}{3}$.
\end{IEEEproof}
\begin{corollary}
  \label{col:r-expectation}
  The average of the normalized correlation matrix elements satisfies:
  \begin{equation*}
    \frac{1}{4^n} \sum_{\bs,\bt \in \lbrace-,+\rbrace^n} \Rho{n}{\bs,\bt} \le
    \Bigl(\frac{2}{3}\Bigr)^n
  \end{equation*}
\end{corollary}
\begin{IEEEproof} The result follows by applying Lemma~\ref{lem:rho-average}
  $n$ times and observing that $\rho_0 = 1$.
\end{IEEEproof}

\section{Rate of Convergence}
Corollary~\ref{col:r-expectation} implies that for large enough $n$, almost all
of non-diagonal entries of $\rho_n$ are small.  However, the bound it gives is
not strong enough to show the asymptotic tightness of the union bound on the
block error probability of polar codes.  For that, one has to show (i) that the
correlations decay like $\bigObig{2^{-(1+\alpha)n}}$ for some $\alpha > 0$, and
(ii) that this bound applies not just to the average value of $\Rho{n}{\bs,\bt}$
but to $\max_{\bt\neq\bs}\Rho{n}{\bs,\bt}$ for the $\bs$'s and $\bt$'s which
index the information channels.

To this end, we establish a probabilistic framework similar to that used in
\cite{arikan09} for proving the channel polarization theorem.

Let $S_1,S_2,\dots,$ be i.i.d $\mathrm{Bernoulli}\left(\frac{1}{2} \right)$
random variables such that $S_i \in \{-,+\}$, define $\bS^n \triangleq
(S_1,S_2,\dots,S_n)$ and $\calF_n \triangleq \sigma(\bS^n)$ as the
$\sigma$-algebra generated by random vector $\bS^n$.  We consider the random
variables $\Z{n}{\bS} = \Ebig{\e{n}{\bS^n} | \bS^n}$ and $\Rho{n}{\bS^n,\bt^n}$
for $\bt^n \in \{-,+\}^n$ which are all $\calF_n$ measurable. 

We show that for any $\alpha > 0$, $\max_{\bt^n \ne \bS^n} \Rho{n}{\bS^n,\bt^n}
\le 2^{-(1+\alpha)n}$ with very high probability for sufficiently large $n$.
\subsection{Closely related $\bs$ and $\bt$}
Let us first focus on $\Rho{n}{\bs,\bt}$ for $\bs$ and $\bt$ sharing a long
common prefix.  Recall that $|\cp{\bs,\bt}|$ denotes the length of this prefix.

\begin{lemma}\label{lem:close_channels}
  Fix $\alpha > 0$. Set $m_n \triangleq 4 \log \bigl(2 (1+\alpha)  n -
  1\bigr)$. Then: 
  \begin{equation*} 
     \lim_{n\to\infty} \Prob{\max_{\bt^n \ne \bS^n: \abs{\cp{\bS^n,\bt^n}} \ge
     m_n} \Rho{n}{\bS^n,\bt^n} \le 2^{-(1+\alpha) n} } = 1
  \end{equation*}
\end{lemma}
\begin{IEEEproof} Let $\bP = \cp{\bS^n, \bt^n}$ and $n_0 = \abs{\bP}$. Observe that
  $\bP$ is a uniformly chosen sign sequence in $\{-,+\}^{n_0}$. According to
  Property~\ref{prop:decreasing}, $\Rho{n_0}{\bP,\bP} = 1$ and: 
  \begin{align*} 
    \Rho{n}{\bS^n,\bt^n} < \Rho{n_0+1}{\bP S_{n_0+1}, \bP t_{n_0+1}}  & = \sqrt{
      \frac{\Z{n_0}{\bP} \Zbar{n_0}{\bP}} {2 + \Z{n_0}{\bP} \Zbar{n_0}{\bP}} }
      \\
    & \le \min \left\{ \sqrt{\frac{1}{2} \Z{n_0}{\bP}}, \sqrt{\frac{1}{2}
    \Zbar{n_0}{\bP}} \right\}.  
  \end{align*}

  Results of \cite{arikan09rate} show that for any fixed $0 < \beta < 1/2$ and
  $\delta > 0$ there exist a $m_0$ such that for $n_0 \ge m_0$   
  \begin{equation*}
    \Prob{\Z{n_0}{\bP} \in [2^{-N_0^{\beta}}, 1 - 2^{-N_0^{\beta}}]} <
    \delta  
  \end{equation*}
  where $N_0 = 2^{n_0}$.

  In particular we take $\beta = \frac{1}{4}$ in the above bound and take $n$
  large enough so that $m_n \ge m_0$. Hence $n_0 \ge m_n \ge m_0$, and with
  probability at least $1-\delta$, $\Z{n_0}{\bP}$ is extremal.  Together with
  $2^{-N_0^{1/4}}\le 2^{-2(1+\alpha)n+1}$ we get\\[1ex]
  \mbox{}\hspace*{\fill} $\displaystyle \Prob{\Rho{n}{\bS^n,\bt^n} \le
  2^{-(1+\alpha)n}} \ge 1 - \delta.  $
\end{IEEEproof}
\subsection{Distantly related $\bs$ and $\bt$}
A more involved task is find and upper-bound on $\Rho{n}{\bs,\bt}$ when $\bs$
and $\bt$ do not have a long common prefix.  For this purpose we first seek an
upper-bound on ${\Rho{n}{\bS^n,\bt^n}}/{\Rho{n-1}{\bS^{n-1},\bt^{n-1}}}$ only in
terms of $\bS^{n-1}$, $S_n$ and $p_n = \abs{ \cp{\bS^n, \bt^n}}$, denoted as
$\chi\left(\bS^{n-1},S_n, p_n\right)$.

To this end, let:
\begin{equation*}
  M\Bigl( S_n, t_n, \Rho{n-1}{\bS^{n-1},\bt^{n-1}}, \Z{n-1}{\bS}, \Z{n-1}{\bt}
  \Bigr)  \triangleq
  \frac{\Rho{n}{\bS^n,\bt^n}}{\Rho{n-1}{\bS^{n-1},\bt^{n-1}}}.  
\end{equation*}

$M\left(s, t, r, a, b\right)$ takes four possible forms according to
\eqref{eq:ratio_cases}, each of which can be bounded as:
\begin{align*}  
  M\left(+,t, r, a, b \right) & \le \min\left\{1, \sqrt{2 a} + r \right\} \\
  M\left(-,t, r, a, b \right) & \le \min\left\{1, \sqrt{2 \bara} + r\right\}
\end{align*}
using Lemma~\ref{lem:F} (and triangle inequality if $s \ne t$):
\begin{lemma}   \label{lem:F}
  Let $f(x) \triangleq \sqfrac{x}{x}$ and $g(x) \triangleq
  \sqfrac{\barx}{x}$. Define $$F(r,a,b) \triangleq 2 f(a) f(b) + g(a) g(b)
  r.$$ Then 
  \begin{equation}
    F(r,a,b) \le \min \left\{1, \sqrt{2 a} + r \right\},
  \end{equation}
  for all $0 \le r \le 1, 0 \le a \le 1, 0 \le b \le 1$.
\end{lemma}
\begin{IEEEproof} Observe that $F(r,a,b) \ge 0$ by construction and:
  \begin{align*}
    F(r,a,b)^2 & = \left(2 f(a) f(b) + g(a) g(b) r \right)^2 \\
    & \stackrel{r < 1}{\le} \left(2 f(a) f(b) + g(a) g(b) \right)^2 \\ 
    & \stackrel{(*)}{\le} \left( 2 f(a)^2 + g(a)^2 \right) \left( 2 f(b)^2 +
    g(b)^2\right)
  \end{align*}
  where (*) follows by Cauchy-Schwarz inequality.
  Furthermore, $2 f(x)^2 + g(x)^2 = \frac{2x}{1+x} + \frac{\barx}{1+x} = 1$
  which proves $F(r,a,b) \le 1$.

  It is also easy to verify $f(x) \le \frac{1}{\sqrt{2}}$ and $g(x) \le 1$ for
  $\forall x \in [0,1]$. Hence:
  \begin{equation*}
    F(r,a,b)  \le \sqrt{2} f(a) + r 
     \le \sqrt{2 a} + r
  \end{equation*}
  where the last inequality follows by observing that $\sqfrac{x}{x} \le
  \sqrt{x}$ since $x \ge 0$.
\end{IEEEproof}

Observe that the upper-bounds on $M$ depend only $\Z{n-1}{\bS}$ and
$\Rho{n-1}{\bS^{n-1},\bt^{n-1}}$. Let us also define
\begin{equation*}
  \Rho{n,p}{\bs^n,*} \triangleq \max_{\substack{ \bt^n \ne \bs^n: \abs{\cp{
  \bs^n,\bt^n}} \le p} }\Rho{n}{\bs^n,\bt^n}.
\end{equation*}
Consequently we may choose:
\begin{subequations}
  \begin{align}
    \chi \left(\bS^{n-1}, +, p_n \right) & = \min \left\{1, \sqrt{2\Z{n-1}{\bS}} +
    \Rho{n-1,p_n}{\bS^{n-1},*} \right\} \label{eq:chi-plus} \\
    \chi \left(\bS^{n-1}, -, p_n \right) & = \min \left\{1, \sqrt{2\Zbar{n-1}{\bS}} +
    \Rho{n-1,p_n}{\bS^{n-1},*} \right\} \label{eq:chi-minus}
  \end{align} 
\end{subequations}

Now we would like to show that $\min_{s_n}\chi\left(\bS^{n-1},s_n,p_n\right)$
gets arbitrarily small with very high probability. For this, we first need the
following lemma:
\begin{lemma} \label{lem:rho-small} 
  For any sequence $p_n$ such that $\lim_{n \to \infty} {\frac{n}{2} - p_n} =
  \infty$ and any fixed $\gamma > 0$,
  \begin{equation}
    \lim_{n \to \infty} \Prob{\forall i \ge \frac{n}{2}: \Rho{i, p_n}{\bS^i,*}
    \le \gamma} = 1.
  \end{equation} 
\end{lemma} 
\begin{IEEEproof} Observe that for fixed $p$, $\Rho{i,p}{\bs^i,*}$ is decreasing
  in $i$ (if $i > p$). Hence $\Rho{n/2,p_n}{\bs^{n/2},*} \le \gamma$ implies
  $\Rho{i,p_n}{\bs^n,*} \le \gamma$ for all $i \ge n/2$.

  Suppose $\bs$ is a sequence such that for some $\bt \ne \bs$ with
  $\abs{\cp{\bs,\bt}} \le p_n$, $\Rho{n/2}{\bs^{n/2},\bt^{n/2}} > \gamma$.
  Recall that $\bs^i$ (resp. $\bt^i$) denotes the subsequence of $\bs$ (resp.
  $\bt$) including its first $i$ elements. 
  
  Define $a_i \triangleq \Rho{i}{\bs^i,\bt^i}$ and $m_i \triangleq a_i /
  a_{i-1}$.  It is clear that $a_{p_n+1} \le \frac{1}{3}$ and $a_i$ is
  decreasing for $i > p_n$ by Properties~\ref{prop:one-third} and
  \ref{prop:decreasing}.
 
  For any $0 < \varepsilon <1$, $a_{n/2} > \gamma$ implies that the number of
  indices $i \in \{p_n+2, p_n+3, \dots, \frac{n}{2}\}$ for which $m_i \le 1 -
  \varepsilon$ is at most $\frac{ \log(3\gamma)}{ \log(1-\varepsilon)}$. 

  Let $l = \frac{n}{2} - p_n - 1$, take $\varepsilon = 1/{\sqrt{l}}$, and
  observe that the number of indices for which $m_i \le 1 - 1/{\sqrt{l}}$ is at
  most 
  \begin{equation*} 
    \frac{\log(3 \gamma)}{\log(1 - 1/\sqrt{l})} \le
    \frac{- \log(3 \gamma)}{1 / \sqrt{l}} = c_\gamma \sqrt{l}, 
  \end{equation*}
  where $c_\gamma$ is a constant that depends on $\gamma$ only.
  These indices partition the interval $[p_n+2 : \frac{n}{2}]$ into at most
  $c_\gamma \sqrt{l}$ segments, one of those must have a length at least
  $c_\gamma^{-1} \sqrt{l}$. Let us only consider this ``long'' segment:

  The fact that $m_i \ge 1 - 1 / \sqrt{l}$ on this segment implies the sign
  sequence $s_{p_n+2},\dots,s_{n/2}$ must be constant on this segment (cf. Proof
  of Lemma~\ref{lem:sample-wise-decrease}). The set of sequences of length $l$
  which have a run of the same sign for an interval of length $c_\gamma^{-1}
  \sqrt{l}$ has probability at most $2 l \cdot 2^{-c_\gamma^{-1} \sqrt{l}}$.
  However, by assumption $l = \frac{n}{2} - p_n -1$ goes to infinity as $n$ gets
  large.  Hence the probability of having such a $\bs$ sequence gets arbitrarily
  small when $n$ gets large. 
\end{IEEEproof}
\begin{lemma} \label{lem:min-chi}
  For any sequence $p_n$ such that $\lim_{n \to \infty} \frac{n}{2} - p_n =
  \infty$ and any fixed $\alpha > 0$
  \begin{equation*}
    \lim_{n \to \infty} \Prob{\forall i > \frac{n}{2}: \min_{s_i} \chi
    \left(\bS^{i-1}, s_i, p_n \right) \le 2^{-4(1+\alpha)} } = 1.
  \end{equation*}
\end{lemma}
\begin{IEEEproof} 
  Let
  \begin{equation*}
    \calG_R(n) \triangleq \left\{\forall i \ge \frac{n}{2}: \Rho{i,p_n}{\bS^i,*}
    \le 2^{-(5+4\alpha)} \right\}.
  \end{equation*}
  Observe that Lemma~\ref{lem:rho-small} implies for any $\delta > 0$ there
  exist a $n_0$ such that $\Prob{\calG_R(n)} \ge 1 - \delta / 2$ for $n \ge
  n_0$.
  
  Let
  \begin{equation*}
    \calG_Z(n) \triangleq \left\{\forall i \ge \frac{n}{2}: \Z{i}{\bS} \notin
    \left[ 2^{-(11+8\alpha)}, 1 - 2^{-(11+8\alpha)} \right] \right\}.
  \end{equation*}
  Likewise, the convergence of $Z$ process implies that there exist a $n_1$
  such that for any $n \ge n_1$ $\Prob{\calG_Z(n)} \ge 1 - {\delta}/{2}$.

  Now \eqref{eq:chi-plus} and \eqref{eq:chi-minus} imply that for $\bS \in
  \calG_R(n) \cap \calG_Z(n)$, $\forall i > \frac{n}{2}$, either
  $\chi\left(\bS^{i-1}, +, p_n \right) \le 2^{-4(1+\alpha)}$ or
  $\chi\left(\bS^{i-1}, -, p_n \right) \le 2^{-4(1+\alpha)}$. For $n \ge
    \max\{n_0,n_1\}$, $\Prob{\calG_R(n) \cap \calG_Z(n)} \ge 1 - \delta$ which
  proves the claim. 
\end{IEEEproof}
\begin{lemma} \label{lem:distant_channels}
  Fix $\alpha > 0$ and let $m_n \triangleq 4 \log \bigl(2 (1+\alpha) n - 1
  \bigr)$ (as in Lemma~\ref{lem:close_channels}).  Then:
  \begin{equation*}
    \lim_{n \to \infty} \Prob{ \max_{\substack{\bt \ne \bS:
      \abs{\cp{\bS,\bt}} < m_n}} \Rho{n}{\bS,\bt} \le 2^{-(1+\alpha)n}} = 1 
  \end{equation*}
\end{lemma}
\begin{IEEEproof} For any $p$, let us define the random variable $B_{n,p}
  \triangleq \Ind{S_n = \arg\min_{s} \chi \left(\bS^{n-1},s,p\right) }$.
  It is easy to see that $\Prob{B_{n,p} = 1 | \calF_{n-1}} = \Prob{B_{n,p} = 0 |
  \calF_{n-1}} = \frac{1}{2}$. 

  Fix $\varepsilon > 0$ and let
  \begin{equation*} 
    \calG_B (n,p,\varepsilon) \triangleq \left\{ \frac{1}{n/2}
    \sum_{i=n/2+1}^{n} B_{i,p} \ge \frac{1-\varepsilon}{2} \right\}.
  \end{equation*}
  Observe that $\Prob{\calG_B(n,p,\varepsilon)}$ is independent of $p$ and 
  by the Weak Law of Large Numbers for any $\delta > 0$ there exist a $n_0$ such
  that $\Prob{\calG_B(n,p,\varepsilon)} \ge 1 - {\delta}/{2}$ for $n \ge
  n_0$.

  Fix $\alpha' > 0$ and define
  \begin{equation*}
    \calG_\chi(n) \triangleq \left\{i > \frac{n}{2}: \min_{s_i}
    \chi(\bS^{i-1}, s_i, m_n) \le 2^{-4(1+\alpha')} \right\}
  \end{equation*}

  Since $\lim_{n \to \infty} \frac{n}{2} - m_n = \infty$, in view of
  Lemma~\ref{lem:min-chi}, there exist $n_1$ such that $\Prob{\calG_\chi(n)} \ge
  1 - {\delta}/{2}$ for $n \ge n_1$. 

  For $n \ge \max\{n_0,n_1\}$, $\Prob{\calG_B(n,m_n,\varepsilon) \cap
  \calG_\chi(n)} \ge 1 - \delta$ and for $\bS^n \in \calG_B(n,m_n,\varepsilon)
  \cap \calG_\chi(n)$ and any $\bt^n \ne \bS^n$ such that
  $\abs{\cp{\bS^n,\bt^n}} < m_n$ we have:
  \begin{align*} 
    \log \left(\Rho{n}{\bS^n,\bt^n}\right) & \le \log
    \left(\Rho{n/2}{\bS^{n/2},\bt^{n/2}} \right) \\   
     & \qquad +  \sum_{i = n/2+1}^{n} \log \left(
    \chi \left(\bS^{i-1}, S_i, m_n \right) \right) \\
    & \stackrel{(*)}{\le} \sum_{i = n/2+1}^n  -4 (1+\alpha')
     B_{i,m_n}\\ 
    & \le - n(1 - \varepsilon)(1+\alpha') .
  \end{align*}
  In the above, (*) follows from the fact that $0 \ge \Rho{n}{\bs,\bt} \le 1$
  and observing that if $B_{i,m_n} = 1$ then $\chi(\bS^{i-1},S_i,m_n) \le 2^{-4
  (1 + \alpha')}$ (as $\bS \in \calG_\chi(n)$), otherwise
  $\chi(\bS^{i-1},S_i,m_n) \le 1$ hence:
  \begin{equation*}
    \log\left( \chi \left( \bS^{i-1}, S_i, m_n \right) \right) \le 
    -4(1+\alpha') B_{i,m_n}.
  \end{equation*}
  For $\bS \in \calG_B(n,m_n, \varepsilon)$, $\sum_{i = n/2+1}^n B_{i,m_n} \ge
  \frac{n(1-\varepsilon)}{4}.$  

  Choosing $\alpha'$ and $\varepsilon$ such that$(1-\varepsilon) (1+\alpha') \ge
  (1+\alpha)$ proves the claim.
\end{IEEEproof}
\begin{theorem} \label{thm:rho_rate}
  For any $\alpha > 0$.
  \begin{equation}
    \lim_{n \to \infty} \Prob{ \max_{\bt \ne \bS} \Rho{n}{\bS,\bt} \le {
      2^{-n(1+\alpha)} }} = 1.
  \end{equation}
\end{theorem}
\begin{IEEEproof} The proof follows by combining the results of
  Lemma~\ref{lem:close_channels} and Lemma~\ref{lem:distant_channels}.
\end{IEEEproof}
\section{Lower Bound on Probability of Error of Polar Codes}
In this section, we use our results on correlations among polarized BECs to
give lower-bounds on block error probability of Polar Codes over BEC.
Recall the analysis of error of the code: The error event $\calE$ is the union
of error events in each of information channels: $ \calE = \bigcup_{\bs \in
\calA} \calE_\bs $ where $\calA \subset \{-,+\}^n$ is the set of information
bits and $\calE_\bs$ denotes the error in $\W{n}{\bs}$.

For a BEC --- with a pessimistic assumption on decoder --- a decision error
happens exactly when an erasure happens. \footnote{A practical decoder can break
  the ties randomly which increases the chance of correctly decoding the bit to
  $\frac{1}{2}$. An analysis analogous to the one we do in this section applies
to such a decoder.} Therefore, $ \calE_\bs = \bigl\{\e{n}{\bs} = 1 \bigr\} $ and
the union bound gives us:
\begin{equation}
  \Prob{\calE} \le \sum_{\bs \in \calA} \Z{n}{\bs}
  \label{eq:upper_bound}
\end{equation}

A trivial lower-bound on the probability of decoding error is obtained by
observing that $\calE \supseteq \calE_\bs$, hence,
$\Prob{\calE} \ge \Prob{\calE_\bs}$ for any $\bs \in \calA$. In particular,
\begin{equation}
  \Prob{\calE} \ge \max_{\bs \in \calA} \Prob{\calE_\bs} = \max_{\bs \in \calA}
  \Z{n}{\bs}.
  \label{eq:trivial_lowerbound}
\end{equation}

However, having the second order statistics, one can use the
inclusion--exclusion principle to obtain a much tighter lower-bound on
probability of error. 

\begin{lemma} \label{lem:lower-bound}
  Let $W$ be a $\BEC{\epsilon}$ and $\calC_n$ be a polar code of
  block-length $N=2^n$ with information bits $\calA_n$. The block error
  probability of such a code, $P_e(\calC_n)$ is lower-bounded as:
  \begin{align}
    P_e(\calC_n)  & \ge  \sum_{\bs\in\calA_n} \Z{n}{\bs}  -  \frac{1}{2} 
      \sum_{\substack{\bs,\bt \in \calA_n:\\ \bs \ne \bt }} \Biggl[\Z{n}{\bs}
      \Z{n}{\bt} \nonumber  \\ & \qquad + \Rho{n}{\bs,\bt} \sqrt{\Z{n}{\bs}
      \Zbar{n}{\bs}} \sqrt{\Z{n}{\bt} \Zbar{n}{\bt}} \Biggr]
    \label{eq:exact_lowerbound}
  \end{align}  
  where $\bZ_n$ vector and $\rho_n$ matrix can be computed via single-step
  recursions explained in Section~\ref{sec:basics}.
 \end{lemma}
\begin{IEEEproof}
  The result follows by applying the inclusion--exclusion principle to
  lower-bound the probability of $\bigcup_{\bs \in \calA_n} \calE_{\bs}$.
\end{IEEEproof}

While the lower-bound given by Lemma~\ref{lem:lower-bound} is already useful in
practice (see Section~\ref{sec:example}), we seek for a lower-bound that is
theoretically more significant.

\begin{theorem} \label{thm:lower-bound}
  Let $W$ be a $\BEC{\epsilon}$ and $R < 1 - \epsilon$. Let $\mathcal{C}_n$ be
  a polar code of block length $N = 2^n$ with information bits $\calA_n$ such
  that $\abs{\calA_n} = \lceil N R \rceil$.  Let $P(N,R,\epsilon)$ be 
  the sum of $\lceil N R \rceil$ smallest elements of the vector $\bZ_n$. Then,
  for any fixed $\delta > 0$ and sufficiently large $n$:
  \begin{equation*} 
    \left(1 - \delta \right) P\left(N,(1-\delta)R,\epsilon\right) \le
    P_e(\calC_n) \le P(N,R,\epsilon).  
  \end{equation*}
\end{theorem}
\begin{IEEEproof}
  The upper-bound is already known and we only need to
  prove the lower-bound. Let 
  \begin{equation*}
    \calD_n = \left\{ \bs \in \{-,+\}^n : \max_{\bt \ne \bs} \Rho{n}{\bs,\bt}
    \le  \delta 2^{-n}\right\}
  \end{equation*}
  By Theorem~\ref{thm:rho_rate} we know that $ \lim_{n \to \infty}
  \frac{\abs{\calD_n}}{N} = 1$.  Let, $\calC'_n$ be the polar code defined by
  the information bits $\calA'_n = \calA_n \cap \calD_n$ and $S_n' \triangleq
  \sum_{\bs \in \calA'_n} \Z{n}{\bs}$.  It is clear that $\lim_{n \to \infty}
  \frac{\abs{\calA'_n}}{\abs{\calA_n}} = 1$, $S'_n \le P(N,R,\epsilon)$ (as
  $\calA_n$ contains $\lceil NR \rceil$ smallest elements of $\bZ_n$), and
  $P_e(\calC'_n) \le P_e(\calC_n)$ as $\calC'_n$ is a sub-code of $\calC_n$. 
  
  Choose $n$ large enough such that $\frac{\abs{\calA'_n}}{\abs{\calA_n}} \ge 1
  - \delta$ and $P(N,R,\epsilon) \le \delta$ (note that this is possible since $R < 1
  - \epsilon$ and the results of \cite{arikan09rate} suggest that
  $P(N,R,\epsilon) = \bigO{2^{-\sqrt{N}}}$). By \eqref{eq:exact_lowerbound}:
  \begin{align*}
    & S'_n - P_e(\calC_n)  \le S'_n - P_e (\calC'_n) \\
    & \quad \le \frac{1}{2}
    \sum_{\substack{\bs,\bt \in \calA'_n: \\ \bs \ne \bt}} \Bigg[ {\Z{n}{\bs}
    \Z{n}{\bt} + \Rho{n}{\bs,\bt} \sqrt{\Z{n}{\bs} \Zbar{n}{\bs}}
    \sqrt{\Z{n}{\bt} \Zbar{n}{\bt}}} \Biggr].
  \end{align*}

  Observe that $\Rho{n}{\bs,\bt} \le \delta / N$ for all $\bs, \bt$ in the above
  summation,  $\sum_{\bs,\bt \in \calA'_n: \bs \ne \bt} \Z{n}{\bs} \Z{n}{\bt}
  \le \sum_{\bs, \bt \in \calA'_n} \Z{n}{\bs} \Z{n}{\bt} = {S'_n}^2$, and
  \begin{align*}
    & \sum_{\bs,\bt \in \calA'_n: \bs \ne \bt} \sqrt{\Z{n}{\bs} \Zbar{n}{\bt}}
    \sqrt{\Z{n}{\bt} \Zbar{n}{\bt}} \\ 
    & \quad \le   \sum_{\bs,\bt \in \calA'_n: \bs \ne \bt} \sqrt{\Z{n}{\bs}}
    \sqrt{\Z{n}{\bt}} 
    \le  \sum_{\bs,\bt \in \calA'_n} \sqrt{\Z{n}{\bs}} \sqrt{\Z{n}{\bt}}  \\ 
    & \quad = \biggl[\sum_{\bs \in \calA'_n} \sqrt{\Z{n}{\bs}} \biggr]^2 
    \stackrel{(*)}{\le} \abs{\calA'_n} \sum_{\bs \in \calA'_n} \Z{n}{\bs}  
    \le N S'_n,
  \end{align*}
  where (*) follows by the Cauchy-Schwarz inequality \footnote{For any set of
    $m$ numbers $x_i, i = 1,2,\cdots,m$: $$\left(\sum_{i = 1}^{m}
    x_i\right)^2 \le m \sum_{i = 1}^{m} x_i^2$$.}.

  Therefore,
  \begin{equation*}
    S'_n - P_e(\calC_n) \le \frac{1}{2} \left[ {S'_n}^2 + \delta S'_n \right] 
    \le \delta S'_n,
  \end{equation*}
  where the last inequality follows by observing that $S'_n \le P(N,R,\epsilon)
  \le \delta$. As a result,
  \begin{equation*}
    (1 - \delta) S'_n \le P_e (\calC_n)
  \end{equation*} 
  $\calC'_n$ is a code of rate $R' \ge (1-\delta)R$ and by
  definition $S'_n \ge P(N,R',\epsilon) \ge P\left(N,(1-\delta) R,
  \epsilon\right)$. Hence we can lower-bound the LHS of the above by
  substituting $S'_n$ with $P\left(N,(1-\delta)R,\epsilon\right)$ which
  completes the proof.
\end{IEEEproof}
\section{Numerical Results} \label{sec:example}
In this section we provide a numerical example which confirms our theoretical
results. We have considered Polar Codes of different rates on a $\BEC{0.5}$ and
computed the upper-bound of \eqref{eq:upper_bound}, the trivial lower-bound of
\eqref{eq:trivial_lowerbound} and the tighter lower-bound of
\eqref{eq:exact_lowerbound}. We emphasize that we have exactly computed the
lower-bound on the error probability by computing the correlation coefficients.
We did the computations for block lengths of $N = 4096$ ($n=12$) and $N=16384$
($n=14$).

\begin{table*}[!htb]
 \pgfplotstableset{%
   columns/rate/.style={%
     column name=$R$,}, %
   columns/sum-Z/.style={%
     column name={$\sum_{\bs \in \calA_n} \Z{n}{\bs}$},},
   columns/max-Z/.style={%
     column name={$\max_{\bs \in \calA_n} \Z{n}{\bs}$ },},
   columns/lower-bound/.style={%
     column name={Lower-bound \eqref{eq:exact_lowerbound}},},
   every head row/.style={before row=\toprule, after row=\midrule},
   every last row/.style={after row=\bottomrule}
 }
 \centering
 \subfloat[$N=4096$] {
   \begin{filecontents*}{0.50-4096.dat}
# Polar Code of block-length 4096 on BEC(0.5)
   rate       sum-Z       max-Z lower-bound
#  0.0029  0.0000e+00  0.0000e+00  0.0000e+00
#  0.0500  4.5722e-82  2.0021e-82  4.5722e-82
#  0.1001  9.5973e-48  2.2679e-48  9.5973e-48
#  0.1501  3.2481e-28  4.2466e-29  3.2481e-28
  0.2002  4.0428e-18  3.4333e-19  4.0428e-18
  0.2502  1.8692e-11  9.2521e-13  1.8692e-11
  0.3000  5.4038e-07  2.2932e-08  5.4036e-07
  0.3501  8.1425e-04  2.1108e-05  8.1237e-04
  0.4001  1.6838e-01  3.4903e-03  1.4438e-01
#  0.4502  6.0020e+00  8.4471e-02 -1.5881e+01
#  0.5000  5.9653e+01  4.9816e-01 -1.8310e+03
\end{filecontents*}
\pgfplotstabletypeset{0.50-4096.dat}
 }

 \subfloat[$N=16384$] {
   \begin{filecontents*}{0.50-16384.dat}
# Polar Code of block-length 16384 on BEC(0.5)
   rate       sum-Z       max-Z lower-bound
#  0.0214  0.0000e+00  0.0000e+00  0.0000e+00
#  0.0500 2.3967e-194 1.2918e-194 2.3967e-194
#  0.1000 4.0607e-102 9.7862e-103 4.0607e-102
#  0.1500  1.0763e-57  7.4788e-59  1.0763e-57
  0.2000  9.3158e-36  4.7150e-37  9.3158e-36
  0.2501  1.3241e-22  3.5356e-24  1.3241e-22
  0.3000  2.3223e-13  5.4016e-15  2.3223e-13
  0.3500  2.6288e-07  3.6113e-09  2.6287e-07
  0.4000  5.4673e-03  4.9056e-05  5.4330e-03
#  0.4500  3.6993e+00  2.3608e-02 -4.4063e+00
#  0.5000  1.6243e+02  4.9977e-01 -1.3496e+04
\end{filecontents*}
\pgfplotstabletypeset{0.50-16384.dat}
 }
 \caption{Bounds on Block Error Probability of Polar Code on 
   $\BEC{0.5}$}
  \label{tab:results}
\end{table*}

As shown in Table~\ref{tab:results}, the proposed lower bound is much tighter
than the trivial one. Moreover, the results show that the lower bound is very
close to the upper bound of \eqref{eq:upper_bound}. This confirms that
$P(N,R,\epsilon)$ (as defined in Theorem~\ref{thm:lower-bound}) is indeed a very
good estimation for block error probability of Polar Codes over BEC.

\bibliographystyle{IEEEtran}
\bibliography{bibliography}
\end{document}